\documentclass[12pt]{article}
\usepackage{a4}
\usepackage{a4wide}

\usepackage{amsmath}
\usepackage{amscd}
\usepackage{amsfonts}
\usepackage{amssymb}
\usepackage{mathrsfs}
\usepackage{fancybox} 
\usepackage{enumerate}

\usepackage{graphicx}
\usepackage{url} 

\usepackage{color}
\usepackage{ulem}

\def\det{\mathrm{det}}

\def\={\stackrel{\bullet}{=}}

\def\({\left(}
\def\){\right)}
\def\[{\left[}
\def\]{\right]}

\def \be {\begin{equation}}
\def \ee {\end{equation}}
\def \beqa {\begin{eqnarray}}
\def \eeqa {\end{eqnarray}}
\def \beal#1 {\begin{align}#1\end{align}}
\def \bes#1 {\begin{equation}\begin{split}#1\end{split}\end{equation}}

\begin{document}

\begin{titlepage}
\title{
\vspace{-2cm}
\begin{flushright}
\normalsize{ 
YITP-20-135
\\ OU-HET-1076}
\end{flushright}
       \vspace{1.5cm}
{\Large Charge Conservation, Entropy Current, and Gravitation}
       \vspace{1.cm}
}
\author{
\large
 Sinya Aoki\thanks{saoki[at]yukawa.kyoto-u.ac.jp },
\; Tetsuya Onogi\thanks{onogi[at]phy.sci.osaka-u.ac.jp},\; 
Shuichi Yokoyama\thanks{shuichi.yokoyama[at]yukawa.kyoto-u.ac.jp},\; 
\\[25pt] 
${}^{*}{}^{\dagger}{}^{\ddagger}$ {\normalsize\it Center for Gravitational Physics,} \\
{\normalsize\it Yukawa Institute for Theoretical Physics, Kyoto University,}\\
{\normalsize\it Kitashirakawa Oiwake-cho, Sakyo-Ku, Kyoto 606-8502, Japan}
\\[10pt]
${}^\dagger$ {\normalsize\it Department of Physics, Osaka University,}\\ 
{\normalsize\it Toyonaka, Osaka 560-0043, Japan,}
}

\date{}

\maketitle

\thispagestyle{empty}

\begin{abstract}
\vspace{0.3cm}
\normalsize

We propose a new class of vector fields to construct a conserved charge 
in a general field theory whose energy momentum tensor is covariantly conserved. 
We show that
there always exists such a vector field in a given field theory even without global symmetry.  
We also argue that
the conserved current constructed from the (asymptotically) time-like vector field can be identified with the entropy current of the system. 
As a piece of evidence we show that the conserved charge defined therefrom satisfies the first law of thermodynamics for an isotropic system
with a suitable definition of temperature. 
We apply our formulation to several gravitational systems such as the expanding universe, Schwarzschild and BTZ black holes, and gravitational plane waves. 
We confirm the conservation of the proposed entropy density under any homogeneous and isotropic expansion of the universe, the precise reproduction of the Bekenstein-Hawking entropy incorporating the first law of thermodynamics, and the existence of gravitational plane wave carrying no charge, respectively.
We also comment on the energy conservation during gravitational collapse in simple models.

\end{abstract}
\end{titlepage}

\section{Introduction}
\label{Intro} 

A central mystery in theory of gravity is that while it is governed by fundamental physics laws, it contains black holes, which behave as thermodynamical objects \cite{Bekenstein:1972tm,Bardeen:1973gs}. 
In particular entropy of a well-known black hole has been shown to be given by the Bekenstein-Hawking formula  \cite{Bekenstein:1973ur,Hawking:1974sw,Bekenstein:1974ax}
(except some cases such as an extremal one \cite{Hawking:1994ii,Teitelboim:1994az,Gibbons:1994ff}): $S=A/4G_N$, where $A$ is the area of the horizon and $G_N$ is the Newton constant. This suggests that information of a black hole such as charges and dynamical degrees of freedom is localized at its surface rather than inside the horizon so as to behave as a membrane-like object \cite{Thorne:1986iy}.

There have been various techniques invented to compute entropy of various types of black holes and to show their thermodynamic relations in and beyond the Einstein gravity 
\cite{Brown:1992bq,Banados:1992gq,Garfinkle:1993xk,Wald:1993nt}.
(See also \cite{Srednicki:1993im,Susskind:1993if,Jacobson:1993vj,Susskind:1994sm,Iyer:1994ys}.) 
These approaches have been developed basically regarding charges including entropy as quasi-local ones evaluated by a surface integral, which enables one to evaluate a charge of black holes without knowing a charge distribution inside the horizon.

In the previous work to holographically realize a black hole with quantum correction \cite{Aoki:2020ztd}, 
the authors of the present paper recognized that the quasi-local energy is not sufficient for precise evaluation of the total energy when matter distributes non-trivially in spacetime. 
We reached a definition to evaluate a total charge of matter by a volume integration of its charge distribution, and proposed a precise definition available on a general curved spacetime with Killing vector fields \cite{Aoki:2020prb}. This definition was recognized in early time \cite{Fock:1959,Trautman:2002zz}, though the validity thereof has not been confirmed by explicit computation. 
We confirmed that this reproduces known results on mass and angular momentum for classic black holes, and that it gives an additional contribution to the known mass formula of any compact star obtained by quasi-local energy\cite{Aoki:2020prb}.  
(See also Ref.~\cite{Angus:2018mep}.)

In this paper, we delve into the proposed definition of a matter charge 
and ask whether it can be extended to the case of geometry without any Killing vector field. 
We conclude that i) a class of vector fields satisfying a particular partial differential equation, which includes the Killing vector fields if they exist,  can make the charge conserved, ii) there always exists such a vector field uniquely for a given initial condition in any field theory with energy momentum tensor covariantly conserved, and iii) the conserved charge constructed from the particular vector field is nothing but the entropy of the system. We present our argument to reach this conclusion with application to several gravitational systems below. 

\section{Charges on a general spacetime}
\label{sec:Proposal}

\subsection{Conservation condition}

Let us consider any field theory on a general curved spacetime. 
We use Greek letters $\mu, \nu, \cdots$ to label an arbitrary fixed coordinate system $x^\mu$, which run from $0$ to $d-1$ with $d$ the dimension of the spacetime.
In the previous paper \cite{Aoki:2020prb} we presented a precise definition of a charge of matter associated with any generic vector field $v^\mu$ as 
\beqa
Q[v](t) &:=& \int_{\Sigma_t} d^{d-1} x\,\sqrt{\vert g\vert} \, T^0{}_\nu v^\nu,
\label{charge}
\eeqa
where $T^\mu\!_\nu$ is the matter energy-momentum tensor given in the system, $\Sigma_t$ is a hypersurface or a time slice at an arbitrarily fixed time $x^0=t$
For instance, if one chooses a vector field as a generator of time translation and that of a space direction with suitable normalization, then the corresponding charges become energy and momentum in that direction, respectively. Superiority of this definition is its manifest general covariance.

It can be shown that the charge defined by \eqref{charge} is conserved or time-independent when $v^\mu$ is a Killing vector field 
and the energy momentum tensor is covariantly conserved
\cite{Fock:1959,Trautman:2002zz,Aoki:2020prb}. 
In this case the defined charge becomes a Noether charge corresponding to global symmetry of the system. 
We emphasize that our proposal can be applied not only to Einstein gravity but also to any other gravitational theories.

A question is whether a charge $Q[v]$ conserves if $v$ is other than a Killing vector field. 
To answer this question we study the time evolution of the charge, which is computed as \cite{Aoki:2020prb}
\beqa
\frac{d Q[v]}{d t} &=& \int_{\Sigma_t} d^{d-1}\vec x \sqrt{\vert g\vert} T^\mu{}_\nu \nabla_\mu v^\nu . 
\eeqa
Therefore if a vector field satisfies the following differential equation
\be
T^\mu{}_\nu \nabla_\mu v^\nu = 0, 
\label{CTE}
\ee 
then it is sufficient for the charge $Q[v]$ to conserve. In this sense we refer to the equation \eqref{CTE} as the conservation condition for the vector $v$. 
In particular any Killing vector field trivially satisfies the conservation condition. This gives our conclusion i) . 

\subsection{Intrinsic vector field}
Since the conservation condition \eqref{CTE} is a 1st order linear partial differential equation, we can convert it to ordinary differential equations by the method of characteristics with a parameter $\tau$. Therefore, for a given energy momentum tensor, there always exists a general solution at least locally in $\tau$, which is determined uniquely once we fix an appropriate initial condition at $\tau=0$ on a hypersurface $\Sigma_{t_0}$ including a choice of a direction of the vector $v$.  

To be more explicit, one can arbitrarily choose a reference vector field $\bar{v}^\mu(x)$
which is defined in the entire spacetime. A simple example is $\bar{v}^\mu(x) := \frac{dx^\mu(\eta)}{d\eta}$ where $\eta$ is a parameter to characterize the evolution of the hypersurface $\Sigma_t$ .
(Furthermore, if we choose $\eta$ to be the global time $x^0$, the reference vector becomes $\bar{v}^\mu(x) = \delta^\mu_0$.)
Then, we can look for the solution of the conservation condition in the form of $v^\mu(x) = c(x) \bar{v}^\mu(x)$ where $c(x)$ is  a scalar function. Eq.\eqref{CTE} reduces to:
\begin{eqnarray}
A^0(x)\partial_0 c(x) +\sum_{\mu\not=0}  A^\mu(x) \partial_\mu c(x) + B(x) c(x) = 0, 
\end{eqnarray}
where $A^\mu(x) := T^\mu{}_\nu (x)\bar{v}^\nu(x)$ and $B(x) := T^\mu{}_\nu(x)\partial_\mu \bar{v}^\nu(x) 
+ T^\mu{}_\nu(x) \Gamma^\mu_{\nu\lambda}\bar{v}^\lambda(x)$.
This can be solved once $c(x^0,\vec x)$ is given at some $x^0$ unless $A^0(x)$ identically vanishes in the entire spacetime
\footnote{In the region where $A^0(x)=0$, one cannot determine the time evolution of $c(x)$ 
. However, since the charge density is proportional to $A^0(x)$ one does not need to know the value of $c(x)$.}.
A choice of  the initial condition  depends on a physical setup of the system. Indeed 
$A^0(x) = T^0{}_\nu (x)\bar v^\nu(x) \not=0$
 is necessary to obtain a non-trivial charge as 
$ Q[v] = \int_{\Sigma_t}\, d^{d-1}\vec x\, \sqrt{\vert g\vert} T^0{}_\nu(x^0,\vec x) \bar{v}^\nu(x^0,\vec x) c(x^0,\vec x)$.

Since the time direction must be chosen so that a component of the stress tensor with time component does not vanish in any reasonable physical system, there always exists a vector field to satisfy \eqref{CTE} proportional to a  time-like vector field (or an asymptotically time-like vector field in the presence of black holes).   We refer to this (asymptotically) time-like vector field as an intrinsic vector field and denote it by $\zeta$. 
\footnote{ The existence of such a vector field for a spherically symmetric gravitational system was pointed out in Ref.~\cite{Kodama:1979vn}.
This vector field may be known as the Kodama vector. We have checked that the Kodama vector satisfies the conservation condition \eqref{CTE}. }
In this paper, we mainly consider a conserved charge associated with $\zeta$, except 
Sec.~\ref{sec:GW} where both conserved energy and momentum are treated.

The above argument establishes our conclusion ii) that there always exists a conserved quantity of the form \eqref{charge} in any field theory on general curved spacetime even without any global symmetry. Thus this conserved charge is different from a Noether charge and must be very special and fundamental for the system.
In the next section, we give physical interpretation for this charge.

\section{Entropy and entropy current}

What is the physical meaning of the conserved charge $Q[\zeta]$ associated with the intrinsic vector field? 
Our answer is {\it entropy} of the system,
\be 
S := Q[\zeta] = \int d^{d-1} x \, s^0, 
\label{entropy}
\ee
where $s^0$ is the time component of an entropy current density defined by
\be 
s^\mu = \sqrt{|g|} T^\mu\!_\nu \zeta^\nu.
\label{entropyCurrent}
\ee
One can easily show that this current density satisfies the ordinary continuity equation $\partial_\mu s^\mu=0$, 
thanks to the conservation condition \eqref{CTE} and the covariant conservation of the energy momentum tensor.

Let us give some remarks for this interpretation.
The entropy is the most fundamental quantity and uniquely defined in
a dynamical system,
and therefore it is physically reasonable that the entropy is related to 
the conserved charge $Q[\zeta]$, which is also uniquely defined for a generic system without global symmetry.
Since the fundamental physics law is expected to be reversible,
it is natural that the entropy of the whole system including matter with gravitational interaction
is conserved in a fundamental theory such as 
general relativity. 
To the contrary, 
if the entropy were not conserved, for example, in a general gravitational system, then
it would mean that there exists extra matter 
producing entropy via unknown interaction (`fifth force'), 
which would be unreasonable.
Further evidence will be presented in following subsections.

\subsection{Analysis for geometry without horizon}
\label{nohorizon}

Let us analyze our proposal in the case where  the coordinate system allows a globally well-defined  unit time evolution  vector field, which we denote by $n^\mu$.
In this case tensors can be decomposed into the longitudinal and transverse components with respect to this vector for each index. 
For example, the energy momentum tensor is decomposed as 
\beqa
T^\mu{}_\nu &=& \rho n^\mu n_\nu + P^\mu{}_\nu -n^\mu J_\nu - J^\mu n_\nu, 
\label{generalMatter}
\eeqa
where 
$\rho := n_\mu T^\mu{}_\nu n^\nu, \
P^\mu{}_\nu := \bar g^\mu{}_\alpha T^\alpha{}_\beta \bar g^\beta{}_\nu,\
J_\nu := n_\alpha T^\alpha{}_\beta \bar g^\beta{}_\nu,$ with $\bar g^\mu{}_\nu =\delta^\mu_\nu +n^\mu n_\nu$, 
and the metric tensor is decomposed into the ADM form as 
\beqa
ds^2 &=& - N^2 (dx^0)^2 + \bar g_{ij} ( dx^i + N^i dx^0) (dx^j + N^j dx^0), ~~~
\eeqa
where $N, N^i$ are called a lapse function and a shift vector, respectively. Note that $J^0=0$
and $n_\mu = -N\delta^0_\mu$.

From the covariant conservation, we have $(\nabla_\mu T^\mu{}_\nu) n^\nu=0$, which boils down to
\beqa
\check\partial \rho + \rho K +P^\mu{}_\nu K^\nu{}_\mu 
+ n^\nu \check\nabla J_\nu - \bar \nabla_\mu J^\mu
&=& 0, 
\label{Conservation2}
\eeqa
where $\check\nabla :=n^\mu\nabla_\mu$, $\bar \nabla_\mu := \bar g^\rho{}_\mu \nabla_\rho$,
$K^\nu{}_\mu := \bar\nabla_\mu n^\nu $ is the extrinsic curvature, and $K := K^\mu{}_\mu$.
Since the intrinsic vector field is defined to be proportional to a time evolution  vector field, we set 
$ \zeta^\mu=-\beta n^\mu$, for which the conservation condition \eqref{CTE}
reduces to
\beqa
\rho \check\partial\beta -\beta P^\mu{}_\nu K^\nu{}_\mu +\beta J_\nu \check\nabla n^\nu -J^\mu \bar \partial_\mu \beta 
&=& 0.
\label{PDE2}
\eeqa
Combining Eqs. (\ref{Conservation2},~\ref{PDE2}) and 
taking a new coordinate $\eta$ satisfying $n^\mu =\displaystyle {\partial x^\mu \over \partial \eta}$,
we obtain
\beqa
\frac{d (\rho\beta)}{d\eta} +\rho\beta K -\bar\nabla_\mu (J^\mu \beta)=0. 
\eeqa
In the case with $J^\mu=0$, this can be easily solved as 
\be 
\beta =\beta_0{\rho_0 \over \rho} \exp \[-\int_{\eta_0}^\eta d\eta K \]
\label{eq:beta}
\ee
where $\rho_0, \beta_0$ are initial values of $\rho, \beta$ at $\eta=\eta_0$, respectively.

The entropy current density becomes
\beqa
s^0 = \sqrt{\bar g}\rho\beta, ~~ 
s^i =-\beta \sqrt{\bar g}( \rho N^i +N J^i)
\label{eq:s0}
\eeqa
where $\bar g:= \det \bar g_{ij}$. 
Thus the continuity equation becomes
\be 
\partial_0 s^0 = \partial_ i(s^0 N^i + N\sqrt{|\bar g|} J^{i} \beta ),
\label{eq:derivative}
\ee
It would be amusing that the shift vector has a physical meaning of the flow velocity of the entropy density when $J^i=0$. 
Note that the entropy density is locally conserved 
if $J^i=N^i=0$.

\subsection{The first law of thermodynamics} 
\label{thermodynamics}

We now show that our entropy current density satisfies the first law of thermodynamics in isotropic systems,
so that the shift vector vanishes and the matter energy momentum tensor is given by a perfect fluid, characterized by $P^\mu{}_\nu = P \bar g^\mu{}_\nu$ and $J_\mu=J^\mu=0$ in \eqref{generalMatter}, and thus $K = \frac{d \log \sqrt{|\bar g|}}{d\eta}$.
In this situation, it follows from the continuity equation \eqref{eq:derivative} that the entropy density itself becomes time-independent. 
Furthermore,
since \eqref{PDE2} in this case becomes
\beqa
\rho\frac{d\beta}{d\eta} &=& \frac{\beta P}{v}  \frac{d v}{d\eta}, 
\eeqa
we can show that $s^0$ in \eqref{eq:s0} satisfies
\beqa
 \frac{d s^0}{d\eta} =\frac{d u}{d\eta}\beta +u \frac{d \beta}{d\eta}= \left(\frac{d u}{d\eta}+ P \frac{d v}{d\eta}\right)\beta,
\label{1stLaw}
\eeqa
where $v:= \sqrt{\vert\bar g\vert}$ is a volume density, and
$u:= \rho v$ is the (internal) energy density corresponding to the energy,
a charge with the unit time evolution vector field as $E:=Q[-n]=\int d^{d-1}x\, u$.
This becomes exactly the first law of thermodynamics if $\beta$ is identified with the inverse temperature, which proves our conclusion iii).
Note that the variation by $\eta$ in eq.~\eqref{1stLaw} is realized by some dynamical process, 
which must satisfy the equation of motion for the matter, as well as the Einstein equation or its variant for gravity if the metric $g_{\mu\nu}$ is dynamical.

Our method determines both the entropy density $s^0$ and the inverse temperature $\beta$ 
for the matter through the gravitational interaction with $g_{\mu\nu}$.
An overall normalizations for both is fixed as an initial condition for the intrinsic vector $\zeta$ in the system, but a ration $s^0/\beta$ is free from such an ambiguity.
Once the normalization is given, the dependence of the temperature on spacetime is completely controlled
by eq.~\eqref{CTE}. This will be seen in Sec. 4.1.

\subsection{Case for black hole geometry}
\label{withhorizon}
Let us consider applying the above formulation to a black hole geometry, where the existence of globally well-defined unit time-like vector field is not guaranteed. Then the analysis done in 
Sec. 3.2
is not generally applicable, and we need to analyze the system by setting the intrinsic vector field $\zeta^\mu = -\zeta \delta^\mu_0$, where $\zeta$ is a function determined to satisfy the conservation condition \eqref{CTE}. 

It happens, however, that the matter energy momentum tensor in black hole geometry vanishes except singularity at the origin \cite{Aoki:2020prb}. 
In such a case, the conservation condition is solved by any time-independent $\zeta$, and there is no further way to determine it.
In this paper, as a provisional prescription, we assume that the energy conservation law holds to satisfy $\frac{1}{\beta}d s^0 = d u + P d v$ as in the previous case, and
we determine $\zeta$ to satisfy the first law of thermodynamics with temperature given by the Hawking temperature identical to the surface gravity normalized by $2\pi$\cite{Hawking:1974sw}. 
We expect that this assumption is in principle verified by setting up a gravitational system such that a smooth geometry with matter gravitationally collapses into the black hole and keeping track of charges during the process. We leave this to future studies.

\section{Applications}
\subsection{Friedmann-Lema\^itre-Robertson-Walker metric}
\label{FLRW}
Let us consider $d$-dimensional Friedmann-Lema\^itre-Robertson-Walker (FLRW) metric, which is given by 
$ds^2 = - (dx^0)^2 + a^2 \tilde g_{ij} dx^i dx^j$, where $a$ is the scale factor dependent only on time, and the Ricci tensor for $\tilde g$ is ${\tilde R}_{ij}=(d-2)k\tilde g_{ij}$ with $k=1,0,-1$ corresponding to a $(d-1)$-dimensional sphere, flat space, hyperbolic space, respectively. 
This is a model of homogeneous and isotropic expanding universe in Einstein gravity with cosmological constant $\Lambda$. In particular, the shift vector vanishes and the energy momentum tensor is given by a perfect fluid, whose density $\rho$ and pressure $P$ are determined by the Einstein or Friedmann equation as
$\rho =\frac{1}{8\pi G_N } \(\frac{ (d-1) (d-2)}2 \frac {k+\dot a^2}{a^2} - \Lambda\right), ~ P=\frac{1}{8\pi G_N } \left((2-d) \big\{ \frac{\ddot a}a + \frac{(d-3) }2 \frac {k+\dot a^2}{a^2} \big\}  + \Lambda\right) , $
where $\dot f = \partial_0 f$.

Since there is a unit time evolution vector field as $n^\mu=-\delta^\mu_0$ and the system is isotropic, this is a case studied in Sec.~3.2, where not only the entropy but also the entropy density are conserved, and the
first law of thermodynamics \eqref{1stLaw} holds. 
The entropy density is computed as $s^0= a^{d-1} \sqrt{\tilde g} \beta \rho $, where $\beta$ is calculated from \eqref{eq:beta} as $u \beta =u_0 \beta_0 =$ constant with $u=\rho \sqrt{\tilde g} a^{d-1}$, since $K=\partial_t \log a^{d-1}$.
Thus in the homogeneous and isotropic expanding universe, the energy density is proportional to the temperature, which decreases during the expansion of the system in any equation of state with non-zero pressure while temperature keeps constant in the case of the pressureless dust,
as evident from $du = - P dv$.
The entropy density satisfies the thermodynamic relation \eqref{1stLaw}. (See also \cite{Kolb:1990vq,BaumannCosmology} and references therein.)
Note that $s^0$ is constant even for the $k=1$ (sphere) case, where the universe first expands, reaches its maximum size, and then contracts. 

\subsection{Schwarzschild black hole} 
\label{Schwarzschild}
We next evaluate the entropy of the Schwarzschild black hole in $d$-dimensional spacetime
with cosmological constant $\Lambda$.
We employ the metric in the
Eddington-Finkelestein coordinate
as 
$
ds^2 = -(1+u(r)) (dx^0)^2 - 2 u dx^0 dr +(1-u(r)) dr^2+ r^2 \tilde g_{ij} dx^i dx^j,
$
where $u(r)=-2\Lambda r^2/(d-2)(d-1)-r_0^{d-3}/r^{d-3}$ and $r_0$ is a constant corresponding to the radius of horizon when $\Lambda=0$.
Note that the metric is non-singular even at $u=\pm 1$, and
a constant $t$ surface is always space-like even inside the horizon for non-negative $\Lambda$. (For negative $\Lambda$, a constant $t$ surface is space-like inside the horizon but becomes time-like for large $r$ satisfying $u(r) > 1$.)
The energy defined by \eqref{charge} 
with a Killing vector $v^\mu=-\delta^\mu_0$ for time translation reproduces the well-known black hole mass $M=V_{d-2}(d-2) r_0^{d-3}/(16\pi G_N)$, where $V_{d-2} :=\int d^{d-2}x \sqrt{\det \tilde g_{ij}}$\cite{Aoki:2020prb}, 
since 
\be 
T^\mu{}_0 = -\rho \delta^\mu_0, ~~ 
\rho :=\frac{d-2}{16\pi G_N}\frac{r_0^{d-3}\delta(r)}{r^{d-2}}.
\label{densityBH}
\ee

This is the case discussed in Sec.~3.3 that the conservation condition is  solved by any time-independent intrinsic vector field because the energy momentum tensor vanishes except $r=0$ as seen from \eqref{densityBH}.
In this situation the intrinsic vector field can be written as $\zeta^\mu = - \zeta(\vec x) \delta^\mu_0$, where $\zeta(\vec x)$ is an arbitrary smooth function of spacial coordinates $\vec x$. Then the entropy current defined by \eqref{entropy} is computed as $s^\mu = \zeta \sqrt g \rho \delta^\mu_0$, where we can take a constant $\zeta=\zeta(\vec 0)$ since $\rho$ is proportional to $\delta(r)$.
Note that the entropy density is localized at the singularity. 
Then the entropy given by \eqref{entropyCurrent} is evaluated as $S=M\zeta$.
As discussed in Sec.~3.3, we determine $\zeta$ to satisfy the first law of thermodynamics $TdS=dM$ identifying $T$ with the Hawking temperature
$T=(d-3 -\frac{2\Lambda r_H^2}{(d-2)})/4\pi r_H$, 
where $r_H$ is the (outer) horizon radius determined by 
$u(r_H)=-1$ \cite{Hawking:1982dh}.
Plugging $S=M\zeta$ into the thermodynamic relation reduces to a differential equation $\zeta + M {d\zeta \over dM} = 1/T$, whose general solution is 
$\zeta = {1 \over M}\int {dM \over T} $. 
Using $dM=\frac{(d-2) V_{d-2} r_H^{d-4} \left((d-3) L^2+(d-1) r_H^2\right)}{16 \pi G_N L^2} dr_H$, 
the integration can be performed as
$
\zeta ={V_{d-2} r_H^{d-2}\over 4 G_N M}. 
$
Remark that we can fix the integration constant since $\zeta\to0$ when $r_0 \to 0$ or $r_H \to 0$, which cannot be fixed only by the thermodynamic relation \cite{Gibbons:1976ue}. 
This reproduces the known result of the blackhole entropy computed by Bekenstein-Hawking formula 
because the area of the horizon is given by $V_{d-2} r_H^{d-2}$. 

\subsection{BTZ black hole} 
\label{BTZ}
Next application is to the BTZ black hole, whose
metric is 
$
ds^2 = -f(r) dt^2 + f(r)^{-1} dr^2 + r^2(d\phi -\omega(r)dt) ^2, 
$
where $f(r)={r^2 \over L^2} - m \theta(r) + {J^2 \over 4r^2} , \omega = {J \over 2 r^2} $. 
It was shown in \cite{Aoki:2020prb} that the energy and the angular momentum defined by using \eqref{charge} choosing $v^\mu= -\delta^\mu_0$ and $v^\mu= \delta^\mu_\phi$ reproduce the well-known result $M={ m \over 8G_N}, ~P_\phi ={J\over 8 G_N }$. 
The matter energy momentum tensor was computed as 
\be 
 T^0{}_\nu ={ \delta(r) \over 16 \pi G_N r}( -{m } \delta_\nu^0 + {J} \delta_\nu^\phi ).
\label{densityBTZBH}
\ee

As discussed in Sec.~3.3 and the case of the Schwarzschild black hole, 
the conservation condition is solved by $\zeta^\mu = - \zeta(\vec x) \delta^\mu_0$, due to \eqref{densityBTZBH}. 
This reduces the entropy density to $s^0 = \zeta m \delta(r) / 16 \pi G_N $, where $\zeta=\zeta(\vec 0)$.
Then the entropy is evaluated as $S=M\zeta$.
As stated in Sec. 3.3 we determine the value of $\zeta$ to satisfy the thermodynamic relation ${\partial S \over \partial M}=1/T$ with $P_\phi$ fixed. 
As in the case of Schwarzschild blackhole, this can be solved as 
$\zeta = {1 \over M}\int {dM \over T} $. 
In the BTZ black hole, the Hawking temperature is given by 
$T={r_+^2 - r_-^2 \over 2\pi L^2 r_+}$ with $r_\pm^2 =m L^2/2 (1 \pm \sqrt{1 -(J/mL)^2})$ \cite{Banados:1992wn}. On the other hand, $dM = {r_+^2 - r_-^2 \over r_+4L^2 G_N} dr_+$ with $P_\phi$ fixed. 
Employing these we can perform the integration as
$\zeta ={2\pi r_+\over 4 G_N M},$
where we fixed the integration constant by $\zeta\to0$ when $r_+ \to 0$. 
This reproduces the known result of the black hole entropy computed by Bekenstein-Hawking formula since the horizon area is given by $2\pi r_+$.
Then the chemical potential for the angular momentum is computed as 
$\mu_\phi = - T {\partial S \over \partial P_\phi}
={ r_- \over r_+L}$, 
since ${\partial S \over \partial P_\phi} 
=- {2\pi r_- L \over r_+^2 - r_-^2} $ with the mass $M$ fixed, which also matches the known result \cite{Banados:1992wn,Carlip:1995qv}. The 1st law of thermodynamics
is given by $TdS = dM-\mu_\phi dP_\phi$.

\subsection{Exact gravitational plane wave}
\label{sec:GW}
Our final example is the gravitational waves, called an exact plane wave solution
discussed in old literature\cite{Weber:1961}.
The metric is given by
$
ds^2 = e^{2\Omega(u)} (dx^2-d\tau^2) + u^2 \left( e^{2\beta(u)} dy^2 + e^{-2\beta(u)}dz^2\right) \notag
$
with $u=\tau - x$.
From the Einstein equation, non-vanishing energy momentum tensors are computed as 
$
T^x{}_x = T^\tau{}_x = -T^\tau{}_\tau = -T^x{}_\tau= \frac{e^{-2\Omega}}{4\pi u}(2\Omega' - u\beta'^2)$ with a prime being a $u$-derivative.
Since the generator of time translation $v_0^\mu=-\delta^\mu_\tau$ and that of $x$-direction $v_x^\mu=\delta^\mu_x$ satisfy the conservation condition, energy and momentum in the $x$ direction are conserved in this system. These are computed from \eqref{charge} as $E =P_x = \frac{V_2}{4\pi} \int dx\, u(2\Omega' - u\beta'^2)$,
where $V_2=\int dydz$.
This implies that the energy and momentum vanish if $2\Omega' - u\beta'^2=0$, in which the matter energy momentum tensor and the Ricci tensor vanish while the Riemann curvature tensor does not when $(u\beta'' +2\beta' -u^2\beta'^3)/u\not=0$.
Thus we observe that there exists a gravitational plane wave which carries neither energy nor momentum, though the Riemann curvature tensor does not vanish everywhere.

If we choose functions $\Omega, \beta$ as $
2\Omega' - u\beta'^2 = \frac{C}{u}\frac{\theta(u+\epsilon) -\theta(u-\epsilon)}{2\epsilon}$
with positive constants $C$ and $\epsilon$,
we obtain $E = P_x = \frac{V_2 C}{4\pi} \int_{\tau-\epsilon}^{\tau+\epsilon} dx\,\frac{1}{2\epsilon} = \frac{V_2 C}{4\pi}.$
The energy momentum tensor is non-zero only on $-\epsilon< \tau-x <\epsilon$, which propagates in the $x$ space at the speed of light in the $\epsilon\rightarrow 0$ limit. 

\section{Discussion}
\label{sec:Discussion}
In this paper we have proposed a procedure to construct a conserved current in any field theory with covariantly conserved energy momentum tensor and interpreted it as the entropy current.
Since any matter generates its energy momentum tensor, which couples to gravitational field, the entropy carried by matter never escapes from the censorship of {\it universal} gravitation. 
This may be a reason why we can find a conserved current in a generic gravitational theory. (See also \cite{Jacobson:1995ab,Verlinde:2010hp}.)
In the applications to black holes, the entropy density is localized at the singularity, while the entropy is still compatible with the Bekenstein-Hawking formula.  
A deeper understanding of this 'information map puzzle' is called for. 

Although the presented formulation may look quite different from known conventional approaches, we confirm that it has successfully reproduced known results with a more unified standpoint. 
In appendix, we also analyze a few simple models for gravitational collapse in this formulation. An advantage may be in that the energy distribution may become meaningful or at least visualizable in such a time-dependent system with event horizon emergent. We draw a picture of energy flow during the gravitational collapses while the total energy is conserved.

Gravitational systems analyzed in this paper are restricted to isotropic and homogeneous system such as the FLRW model and a few classic blackholes whose matter distribute only at the singularity. In these systems, in order to solve the conservation condition it is sufficient to choose the intrinsic vector field to be proportional to the time evolution vector field. However in a more general system, there needs to be more vector fields to characterize the system. 
In such a situation, 
the intrinsic vector field  must be expanded by all such vector fields in general. As a result the simple thermodynamic relation for the entropy in Sec.~\ref{thermodynamics} obtained in this paper will be modified significantly. It would be very interesting to find out the corresponding dynamical thermodynamic relation and the local inverse temperature in such a more general situation. 

The application of the method developed in this paper is also open to any effective field theory on a fixed background metric. This application may shed new light to thermodynamics and hydrodynamics. (See also \cite{Sasa:2015zga}.)

While the entropy is always conserved in situations considered in this paper, there are several ways to violate the conservation of the entropy: by changing the boundary conditions of fields or by violating the covariant conservation of energy momentum tensor.
In the latter case, while the effect may be compensated by finding a vector field $v$ to satisfy $\nabla_\mu (T^\mu{}_\nu v^\nu) = 0$, 
the original conservation condition may be suitable to investigate the second law of thermodynamic.

We leave these interesting issues to future studies.

\section*{Acknowledgement}
This work is supported in part by the Grant-in-Aid of the Japanese Ministry of Education, Sciences and Technology, Sports and Culture (MEXT) for Scientific Research (Nos.~JP16H03978, JP18K03620, JP18H05236, JP19K03847). 
T. O. would like to thank YITP for their kind hospitality during his
stay for the sabbatical leave from his home institute.

\appendix

\section{Local energy density in gravitational collapse and energy conservation}
In this appendix,  simple gravitational collapses~\cite{Adler:2005vn}, extended to an arbitrary dimensions with non-zero cosmological constant, are analyzed in terms of conserved charge proposed in this paper, which turns out to be the total energy of the system in these cases.
We will see how energy distribution varies during the collapse while the total energy is conserved even after a black hole formation.

\subsection{Thick Light shells}
\label{sec:Shell}
\begin{figure}[bth]
\begin{center}
\includegraphics[width=0.35\hsize,angle=270]{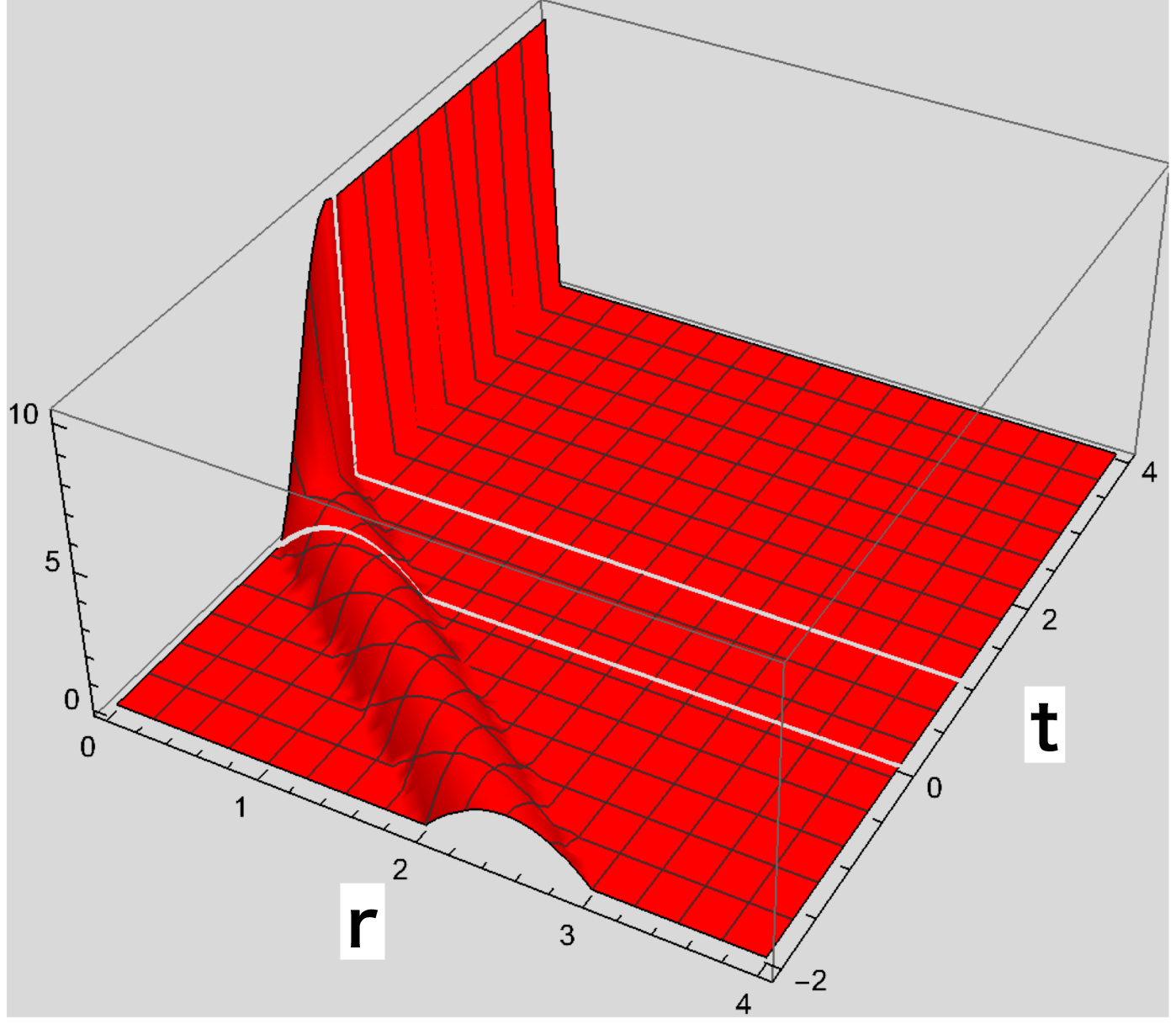}
\includegraphics[width=0.35\hsize,angle=270]{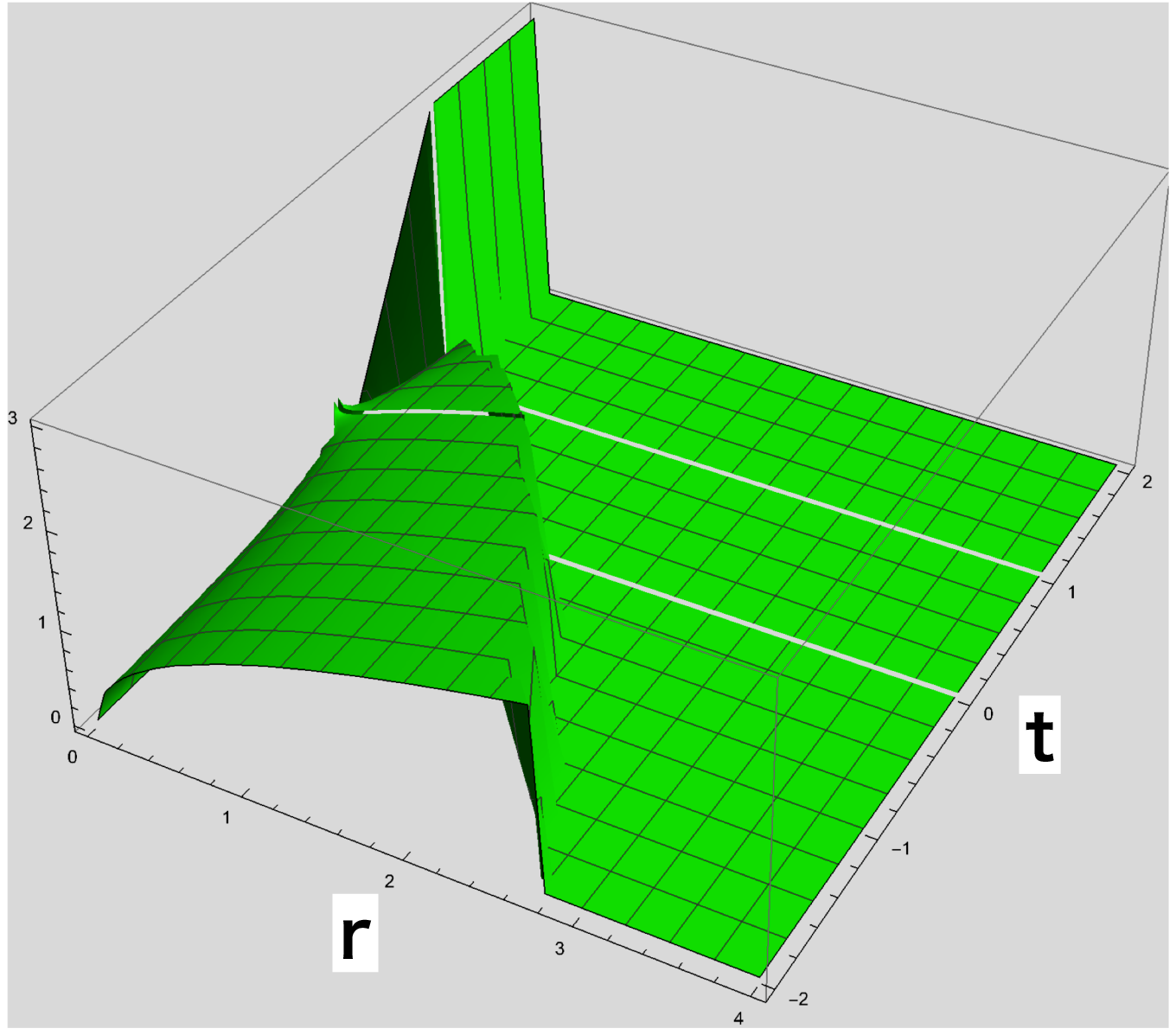}
\end{center}
\caption{Energy density during the gravitational collapse as a function of $r$ and $t$. The black hole formation at $r=0$ starts at $t=0$ and ends at $t=1$.
The $\delta$ function at the origin ($r=0$) for the black hole is approximated by the step function for visibility. 
(Left) Thick light shells in the Eddington-Finkelstein coordinate.
(Right) Zero pressure fluid in the Painleve-Gullstrand coordinate.
}
\label{fig:Collapse}
\end{figure}

We first consider a simple model of gravitational collapses 
for thick light shells~\cite{Adler:2005vn},
whose metric 
in the Eddington-Finkelstein coordinate is given by
\beqa
g_{\mu\nu} dx^\mu dx^\nu &=&- (1+u) dt^2 - 2u dt dr+(1-u)dr^2 + r^2 \bar g_{ij} dx^i dx^j,
\eeqa
where $\bar g_{ij}$ is the $d-2$ metric for a sphere,  
\beqa
u(r,t) &=& \delta u(r,t) -\frac{\Lambda r^2}{(d-2)(d-1)} ,\quad
\delta u( r,t) := -\frac{m(r,t)}{r^{d-3}} ,
\eeqa
$m(r,t) = M\theta(r)$, $M \theta(r)F({t+r\over \Delta})$, $0$ at
$t+r> \Delta$, $0 \le t+r\le \Delta$, $t+r < \Delta$, respectively,
with a monotonically increasing function $F(x)$ satisfying $F(0)=0$ and $F(1)=1$.
Thus the energy momentum tensor is evaluated as
\beqa
T^t{}_t &=& \frac{(d-2)}{16\pi G_N} \frac{\left( r^{d-3} \delta u\right)_r}{r^{d-2}},\quad
T^r{}_r =  \frac{(d-2)}{16\pi G_N}\left[ \frac{ \left( r^{d-3} \delta u\right)_r}{r^{d-2}}
-\frac{2 (\delta u)_t}{r}\right],  \\
 T^t{}_r &=& \frac{(d-2)}{16\pi G_N} \frac{ (\delta u)_t}{r}= -  T^r{}_t,   \quad
 T^i{}_j  = \frac{ \delta^i_j}{16\pi G_N }
 \left[
 \frac{ ( r^{d-3}\delta )_{rr}}{r^{d-3}}  -\frac{ 2 ( r^{d-3}\delta )_{rt}}{r^{d-3}} +(\delta u)_{tt} 
 \right] ,
\eeqa
and $\zeta^\mu= -\delta^\mu_t$ satisfies  eq.~\eqref{CTE} for the intrinsic vector. 
The corresponding charge (energy) is conserved for all $t$ as
\beqa
E(t) &=& -\int d^{d-1} x\, \sqrt{\vert g\vert} \, T^t{}_t =  \frac{(d-2) V_{d-2}}{16\pi G_N} M,
\eeqa
and Fig.~\ref{fig:Collapse} (Left) show a distributions of local energy $-\sqrt{\vert g\vert} T^t{}_t$
with $F(x)=3 x^2 -2 x^3$.

\subsection{Zero pressure fluid (dust) }
\label{sec:Dust}
Zero pressure fluid is handled by the Painleve-Gullstrand coordinate as~\cite{Adler:2005vn}
\beqa
g_{\mu\nu} dx^\mu dx^\nu &=& (\psi^2-1) dt^2 -2\psi dt dr + dr^2 +r^2 \bar g_{ij} dx^i dx^j, 
\eeqa
where
$\psi(r,t) =-\sqrt{u(r,1)}$, $-\sqrt{u(r,\lambda)}$ at $t-\Delta +G(r,M) >0$, $t-\Delta +G(r,M)\le 0$, 
\beqa
u(r,\lambda) =\frac{M\theta(r)\lambda}{r^{d-3}} + \frac{2\Lambda r^2}{(d-2)(d-1)}, \quad
G(r,M) = \int_0^r\frac{ds}{\sqrt{u(s,1)}}, 
\eeqa
and $0\le \lambda \le \Delta$ is determined as a function of $r,t$ by
$t -h(\lambda) + G(r,M\lambda) = 0$
through a given monotonically increasing function of $h(\lambda)$ with $h(0)=0$ and $h(1)=\Delta$.

As in the previous case, $\zeta^\mu = -\delta^\mu_t$ becomes the intrinsic vector,
whose corresponding charge (energy) is conserved for all $t$.
Fig.~\ref{fig:Collapse} (Right) shows a distribution of energy density, given by
$\sqrt{\vert g\vert}  \frac{d-2}{16\pi G_N} \frac{(r^{d-3}\delta\psi^2)_r}{r^{d-2}}$
with $\delta\psi^2 := \psi^2 -\frac{2\Lambda r^2}{(d-2)(d-1)} $ for $h(\lambda)=\Delta \lambda$ and $\Lambda=0$.

\subsection{Matters with non-zero pressure }
\label{sec:Matter}
As a final example, 
let us consider a gravitational collapse of matter with non-zero pressure,
whose metric in the Vaidya coordinate is given by
\beqa
g_{\mu\nu} dx^\mu dx^\nu &=& -f(r,v) dv^2 + 2dv dr + r^2 \bar g_{ij} dx^i dx^j, ~~~
\eeqa
where $f(r,v) = 1 +\delta f -\frac{2\Lambda r^2}{(d-2)(d-1)} $ with $f(r,v) = 1 +\delta f -\frac{2\Lambda r^2}{(d-2)(d-1)} $, and $v=t+r$ with $t$ in the Eddington-Finkelstein coordinate~\cite{Adler:2005vn}.
The EMT becomes 
\beqa
T^v{}_v&=&T^r{}_r = \frac{d-2}{16\pi G_N} \frac{(r^{d-3} \delta f)_r}{r^{d-2}}, \
T^r{}_v= -\frac{d-2}{16\pi G_N} \frac{(\delta f)_v }{r},\
T^i{}_j =\frac{\delta^i_j}{16\pi G_N} \frac{(r^{d-3} \delta f)_{rr}}{r^{d-3}}, ~~~
\eeqa
and  $T^v{}_r=0$,
which describes matters with non-zero pressure (non-perfect fluid)
as well as the energy momentum transfer due to $T^r{}_v$. 
Matters with non-zero pressure indeed describes  infalling massive matters, since $v+r = t+ 2r =\mbox{constant}$ represents propagations slower than light.   
It is easy to see that  $\zeta^\mu = -\delta^\mu_v$ is the intrinsic vector, so that
the corresponding charge (energy) is conserved for all $t$.

\bibliographystyle{utphys}
\bibliography{Entropy}

\end{document}